\documentclass[aps,prb,twocolumn,groupedaddress,amsmath,amssymb,superscriptaddress]{revtex4-2}

\usepackage{graphicx}
\usepackage{dcolumn}
\usepackage{bm}
\usepackage[none]{hyphenat}
\usepackage[final]{changes}
\begin{document}


\title{Tunable Luttinger liquid and correlated insulating states in one-dimensional moiré superlattices}

\author{Jiajun Chen}
\thanks{These authors contribute equally to this work.}
\affiliation{Key Laboratory of Artificial Structures and Quantum Control (Ministry of Education), School of Physics and Astronomy $\&$ Tsung-Dao Lee Institute, Shanghai Jiao Tong University, Shanghai 200240, China.}

\author{Bosai Lyu$^{*}$}
\email{bslyu23@sjtu.edu.cn}
\affiliation{Key Laboratory of Artificial Structures and Quantum Control (Ministry of Education), School of Physics and Astronomy $\&$ Tsung-Dao Lee Institute, Shanghai Jiao Tong University, Shanghai 200240, China.}

\author{Liguo Wang}
\thanks{These authors contribute equally to this work.}
\affiliation{Key Laboratory of Artificial Structures and Quantum Control (Ministry of Education), School of Physics and Astronomy $\&$ Tsung-Dao Lee Institute, Shanghai Jiao Tong University, Shanghai 200240, China.}

\author{Shuo Lou}
\affiliation{Key Laboratory of Artificial Structures and Quantum Control (Ministry of Education), School of Physics and Astronomy $\&$ Tsung-Dao Lee Institute, Shanghai Jiao Tong University, Shanghai 200240, China.}

\author{Xianliang Zhou}
\affiliation{Key Laboratory of Artificial Structures and Quantum Control (Ministry of Education), School of Physics and Astronomy $\&$ Tsung-Dao Lee Institute, Shanghai Jiao Tong University, Shanghai 200240, China.}

\author{Tongyao Wu}
\affiliation{Key Laboratory of Artificial Structures and Quantum Control (Ministry of Education), School of Physics and Astronomy $\&$ Tsung-Dao Lee Institute, Shanghai Jiao Tong University, Shanghai 200240, China.}

\author{Yi Chen}
\affiliation{Key Laboratory of Artificial Structures and Quantum Control (Ministry of Education), School of Physics and Astronomy $\&$ Tsung-Dao Lee Institute, Shanghai Jiao Tong University, Shanghai 200240, China.}

\author{Cheng Hu}
\affiliation{Key Laboratory of Artificial Structures and Quantum Control (Ministry of Education), School of Physics and Astronomy $\&$ Tsung-Dao Lee Institute, Shanghai Jiao Tong University, Shanghai 200240, China.}

\author{Guibai Xie}
\affiliation{Key Laboratory of Artificial Structures and Quantum Control (Ministry of Education), School of Physics and Astronomy $\&$ Tsung-Dao Lee Institute, Shanghai Jiao Tong University, Shanghai 200240, China.}

\author{Kenji Watanabe}
\affiliation{Research Center for Electronic and Optical Materials, National Institute for Materials Science, 1-1 Namiki, Tsukuba 305-0044, Japan.}

\author{Takashi Taniguchi}
\affiliation{Research Center for Materials Nanoarchitectonics, National Institute for Materials Science, 1-1 Namiki, Tsukuba 305-0044, Japan.}

\author{Mengzhou Liao}
\affiliation{College of Materials Science and Engineering, Sichuan University, Chengdu, 610065, China}

\author{Wei Yang}
\affiliation{Beijing National Laboratory for Condensed Matter Physics, Key Laboratory for Nanoscale Physics and Devices, Institute of Physics, Chinese Academy of Sciences, Beijing 100190, China}
\affiliation{Songshan Lake Materials Laboratory, Dongguan 523808, China}

\author{Guangyu zhang}
\affiliation{Beijing National Laboratory for Condensed Matter Physics, Key Laboratory for Nanoscale Physics and Devices, Institute of Physics, Chinese Academy of Sciences, Beijing 100190, China}
\affiliation{Songshan Lake Materials Laboratory, Dongguan 523808, China}

\author{Binbin Wei}
\affiliation{Institute of System Engineering, Beijing 100191, China}

\author{Xiaoqun Wang}
\affiliation{School of Physics and Institute for Advanced Studies of Physics, Zhejiang University, Hangzhou 310058, Zhejiang, China}

\author{Qi Liang}
\affiliation{Key Laboratory of Artificial Structures and Quantum Control (Ministry of Education), School of Physics and Astronomy $\&$ Tsung-Dao Lee Institute, Shanghai Jiao Tong University, Shanghai 200240, China.}
\affiliation{Collaborative Innovation Centre of Advanced Microstructures, Nanjing University, Nanjing 210093, China.}

\author{Guohua Wang}
\affiliation{Key Laboratory of Artificial Structures and Quantum Control (Ministry of Education), School of Physics and Astronomy $\&$ Tsung-Dao Lee Institute, Shanghai Jiao Tong University, Shanghai 200240, China.}
\affiliation{Collaborative Innovation Centre of Advanced Microstructures, Nanjing University, Nanjing 210093, China.}

\author{Jie Ma}
\affiliation{Key Laboratory of Artificial Structures and Quantum Control (Ministry of Education), School of Physics and Astronomy $\&$ Tsung-Dao Lee Institute, Shanghai Jiao Tong University, Shanghai 200240, China.}
\affiliation{Collaborative Innovation Centre of Advanced Microstructures, Nanjing University, Nanjing 210093, China.}

\author{Dong Qian}
\affiliation{Key Laboratory of Artificial Structures and Quantum Control (Ministry of Education), School of Physics and Astronomy $\&$ Tsung-Dao Lee Institute, Shanghai Jiao Tong University, Shanghai 200240, China.}
\affiliation{Collaborative Innovation Centre of Advanced Microstructures, Nanjing University, Nanjing 210093, China.}

\author{Guorui Chen}
\affiliation{Key Laboratory of Artificial Structures and Quantum Control (Ministry of Education), School of Physics and Astronomy $\&$ Tsung-Dao Lee Institute, Shanghai Jiao Tong University, Shanghai 200240, China.}
\affiliation{Collaborative Innovation Centre of Advanced Microstructures, Nanjing University, Nanjing 210093, China.}

\author{Tingxin Li}
\affiliation{Key Laboratory of Artificial Structures and Quantum Control (Ministry of Education), School of Physics and Astronomy $\&$ Tsung-Dao Lee Institute, Shanghai Jiao Tong University, Shanghai 200240, China.}
\affiliation{Collaborative Innovation Centre of Advanced Microstructures, Nanjing University, Nanjing 210093, China.}

\author{Mingpu Qin}
\email{qinmingpu@sjtu.edu.cn}
\affiliation{Key Laboratory of Artificial Structures and Quantum Control (Ministry of Education), School of Physics and Astronomy $\&$ Tsung-Dao Lee Institute, Shanghai Jiao Tong University, Shanghai 200240, China.}
\affiliation{Hefei National Laboratory, University of Science and Technology of China, Hefei 230088, China}

\author{Xiao Yan Xu}
\email{xiaoyanxu@sjtu.edu.cn}
\affiliation{Key Laboratory of Artificial Structures and Quantum Control (Ministry of Education), School of Physics and Astronomy $\&$ Tsung-Dao Lee Institute, Shanghai Jiao Tong University, Shanghai 200240, China.}
\affiliation{Hefei National Laboratory, University of Science and Technology of China, Hefei 230088, China}

\author{Zhiwen Shi}
\email{zwshi@sjtu.edu.cn}
\affiliation{Key Laboratory of Artificial Structures and Quantum Control (Ministry of Education), School of Physics and Astronomy $\&$ Tsung-Dao Lee Institute, Shanghai Jiao Tong University, Shanghai 200240, China.}
\affiliation{Collaborative Innovation Centre of Advanced Microstructures, Nanjing University, Nanjing 210093, China.}



\begin{abstract}
Two-dimensional moiré superlattices have been extensively studied, and a variety of correlated phenomena have been observed. However, their lower-dimensional counterpart, one-dimensional (1D) moiré superlattices, remain largely unexplored. Electrons in 1D are generally described by Luttinger liquid theory, with universal scaling relations depending only on the Luttinger parameter g. In particular, at half-filling, Umklapp scattering plays a crucial role, as it can significantly change the conductance-temperature scaling relation and lead to Mott insulators. However, this prediction has never been observed since doping an empty band to half-filling was extremely difficult. Here, we show that the marriage of moiré superlattices and 1D electrons makes it possible to study the Luttinger liquid in an exceptionally wide filling region simply by electrical gating. We perform transport measurements on 1D moiré superlattices of carbon nanotubes on hexagonal boron nitride (hBN) substrates, and observe correlated insulating states at 1/4 and 1/2 fillings of the superlattice mini-band, where Umklapp scattering becomes dominant. We also observe a T-linear conductance at these commensurate fillings over a range of temperatures. Strikingly, the T-linear conductance leads to a strongly suppressed Luttinger parameter, suggesting a state of extreme correlation.
\end{abstract}

\maketitle


\section{Introduction}
Two-dimensional (2D) van der Waals superlattices have exhibited numerous correlated electronic phases\cite{1,2,3,4,5,6,7,8,9,10,11,12,13}, including Mott insulator\cite{2,4}, superconductivity\cite{1,3}, Chern insulators\cite{5,14}, ferromagnetism\cite{5,11,15}, charge density wave (CDW)\cite{14}, and Wigner crysta\cite{8,16}. Basically, the emergence of the above correlated phases can be described by the Hubbard model with $U>W$, where $U$ represents the Coulomb repulsion energy and $W$ the electron kinetic energy\cite{2,6,17}. To increase the correlation strength and observe those correlated states, two main approaches have been widely employed: one is creating a flat band with a small width $W$ by stacking 2D materials as in the magic-angle bilayer graphene\cite{1,2}, ABC trilayer graphene/hexagonal boron nitride (hBN) superlattices\cite{3,4,5}, and twisted double bilayer graphene\cite{7,9}; the other one is altering the Coulomb energy $U$ via tuning the Coulomb screening\cite{6,17}. So far, those correlated phenomena have mostly been observed in 2D superlattices. 

In contrast to 2D systems, electrons in one-dimensional (1D) materials are naturally interacting to each other more strongly due to the reduced electrostatic screening. In addition, 1D electrons have larger quantum fluctuation. As a result, quasi-free electrons within Fermi liquid theory break down, and electrons confined in 1D behave as a Luttinger liquid, a strongly correlated electronic matter\cite{18,19,20,21,22}. Many exotic phenomena predicted by Luttinger liquid theory, such as spin-charge separation\cite{23,24,25,26} and power-law dependence of spectral functions near the Fermi level\cite{27,28}, has been successfully observed in 1D systems. Carbon nanotube (CNT) is one of the best-known 1D materials. Structurally, CNTs have a tiny diameter of $\sim$\;1 nm and an ultralong length up to centimeters; electronically, charge carriers in CNTs possess exceptionally high mobility\cite{29,30,31} of up to 100,000\;$\mathrm{cm}^2/\mathrm{V}\cdot\mathrm{s}$. It has been predicted that when a CNT’s electronic band is half-filled, Umklapp scattering dominates and a metal-insulator phase transition occurs\cite{18,19,21}. However, this theoretical prediction has never been realized because it is experimentally challenging to achieve the half-filled band in a bare CNT. One possible solution to this challenge is to take advantage of CNT moiré superlattices with mini electronic bands that enables achieving half-filling at a relatively low doping concentration\cite{32}.

In this letter, we created 1D CNT superlattices with a moiré period of 14 nm through direct growth of carbon nanotubes (CNTs) on an atomically flat hBN substrate. We further performed transport measurement on the CNT superlattices, and observed correlated insulating states at 1/4 and 1/2 fillings of the superlattice hole mini-band, corresponding to 1 and 2 holes per unit supercell, respectively. We ascribe the correlated insulating state at 1/2 filling to Umklapp scattering\cite{18,27,28}, while the correlated insulating state at 1/4 filling requires more caution. It may be due to the combination of Umklapp scattering and polarization of the valley degree of freedom. At these specific fillings, the Luttinger liquid theory is thought to be still valid over a range of temperatures above the correlated gap\cite{18}. However, when we fit the transport behavior with the power law, we get T-linear conductance, leading to strongly suppressed Luttinger parameter, indicating an extremely strong electron correlation. Moreover, the filling factor can be continuously tuned via gating, so the relevant scattering process can evolve from the forward channel to the Umklapp channel\cite{18}, leading to a largely tunable Luttinger parameter. Our observations demonstrate the 1D moiré superlattices to be an ideal platform for studying Luttinger liquid with extremely tunable Luttinger parameter and exploring novel correlated electron phases.
\begin{figure*}
\includegraphics[width=15cm]{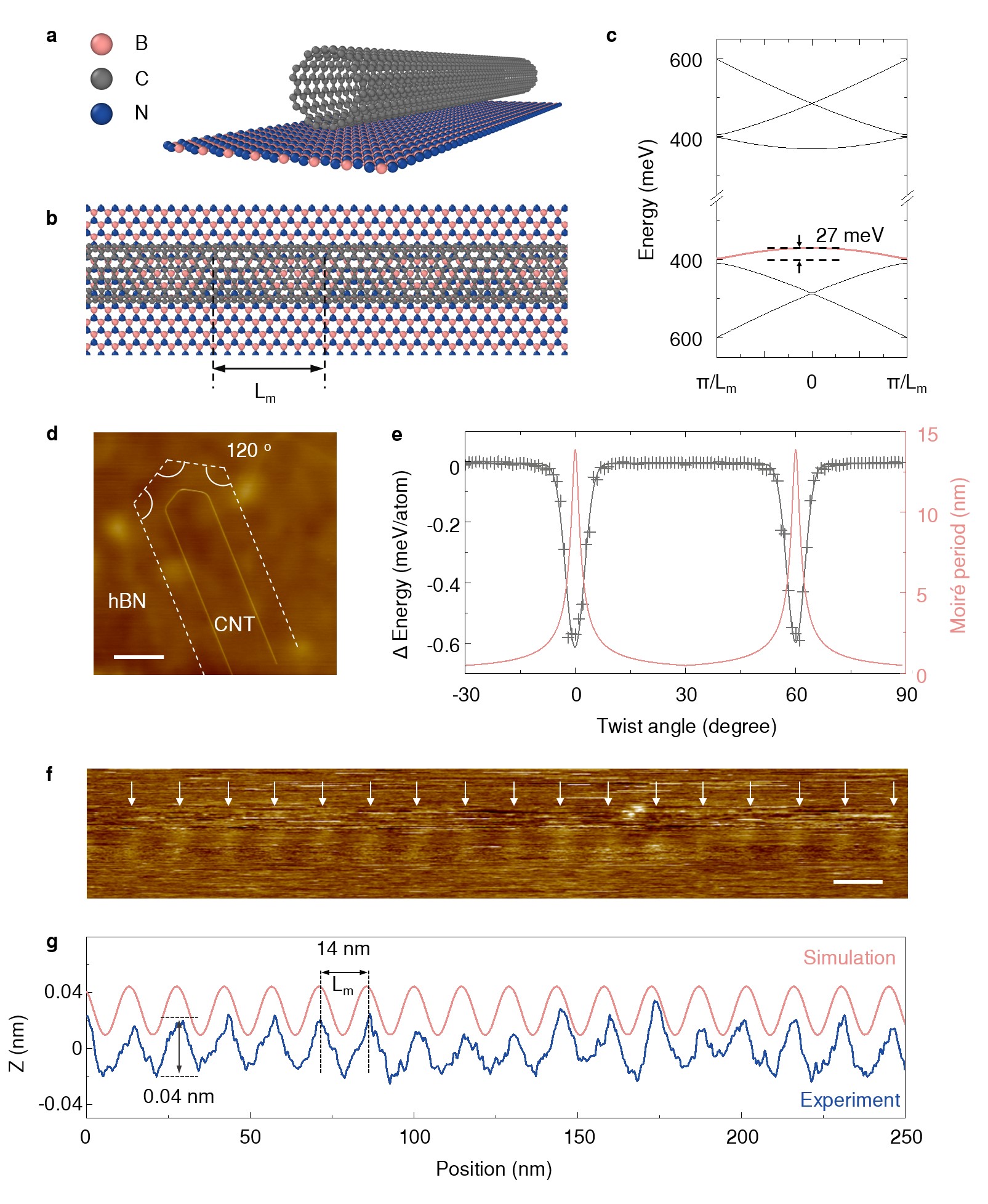}
\centering
\caption{\label{fig1} 1D moiré superlattices of CNTs on hBN. (a) Schematic of a CNT/hBN heterostructure. (b) Top view of the CNT/hBN heterostructure, showing a clear 1D moiré pattern. (c) Band structure of a (10,0) CNT on hBN substrate. New bandgaps are opened by the moiré potential, generating narrow minibands. (d) An as-grown CNT on hBN substrate. All bending angles are 120°, implying a perfect alignment with the hBN substrate. Scale bar: 1 $\mu$m. (e) Dependence of the CNT/hBN interlayer stacking energy (black curve) and the moiré period (red) on the twist angle. At zero twist angle the stacking energy is minimal and the moiré period is maximal $\sim$\;14 nm. (f) A zoom-in AFM topography image exhibits a well-defined periodic pattern along the CNT. Arrows indicate each local maximum in height along the CNT. Scale bar: 15 nm. (g) Experimental height profile (blue) extracted from the AFM image in (f), showing a period of $\sim$\;14 nm and a corrugation of $\sim$\;0.04 nm. The orange line is the MD simulated height profile. }
\end{figure*}

\section{1D moiré superlattices of CNTs on hBN}
Figure 1a presents a schematic structure of the CNT/hBN heterostructure. Well-defined 1D moiré superlattice due to the lattice mismatch can be seen in Fig.\;1b. The electronic bands of CNTs are modified by external periodic potential from hBN, generating superlattice minibands with largely reduced bandwidth\cite{33}. Figure 1c displays the band structure of a CNT with chiral indices of (10, 0). A series of bandgaps are opened by the moiré potential at the boundary of the superlattice Brillouin zone, resulting in isolated narrow bands with minimum bandwidth of only 27 meV (red curve in Fig. 1c). Noted that such a bandwidth (kinetic energy) is smaller than the estimated on-site Coulomb repulsion energy $U=\frac{e^2}{4\pi\varepsilon\varepsilon_0L_m}\approx50\mathrm{~meV}$ ($\varepsilon\approx2.1$, the averaged dielectric constant of the environment). Furthermore, it is expected that the electrostatic screening from the material itself is largely reduced in 1D materials, resulting in even stronger Coulomb interaction. Therefore, novel strong correlation phenomena are likely to emerge in the 1D superlattices.

The 1D CNT/hBN moiré superlattice is synthesized through direct growth of single-walled CNTs on atomically flat hBN flakes using chemical vapor deposition (CVD) (see method for the growth details)\cite{34}. Strikingly, most of the as-grown CNTs on hBN exhibit polylines with consistent bending angles of either 120° or 60°, as shown in the atomic force microscopic (AFM) images in Fig.\,1d and Fig.\,S1. Such a peculiar bending angle indicates that the CNT is crystallographically aligned (zero twist angle) with the hBN substrate, as the hBN crystal has three-fold rotational symmetry. The self-aligning behavior can be understood by considering that the stacking energy of CNT on hBN is significantly reduced (by 0.6 meV/atom) for the perfect alignment configuration, as indicated by our molecular dynamics (MD) simulations (Fig.1e.\;See more details in SI section 2). The aligned CNTs on hBN with zero twist angle are expected to display moiré pattern with a period of approximately 14 nm, due to the 1.8\% lattice mismatch between hBN and graphene. A fine scan along a well-aligned nanotube indeed reveals a periodic oscillation in its height (Fig.\,1f, Figs.\,S5a and S6b). Each oscillation peak in Fig.\,1f, denoted by a white arrow, represents a moiré superlattice period of approximately 14 nm. This period is consistent with previous observations in perfectly aligned graphene on hBN substrate\cite{35}. For CNTs with a small stacking twist angle of 1.1°, the moiré period decreases to 9.6 nm (Figs.\,S4-5). Increasing the twist angle further leads to a decrease in both the moiré period and the moiré modulation, eventually rendering them undetectable (Figs.\,S2 and S6). In the case of perfect alignment (zero twist angle), 1D moiré patterns generally exist in all CNTs with different chiralities, although the moiré period may slightly vary, ranging from 12 nm to 16 nm (See SI section 4). The height corrugation in the 1D moiré patterns is $\sim$\;0.04 nm (Fig.\,1g, blue curve), which is consistent with our MD simulations (orange curve) and comparable to that of 2D graphene on hBN. Such a periodic modulation in the CNT structure will probably induce an observable change in its electronic property.

\begin{figure*}
\includegraphics[width=17cm]{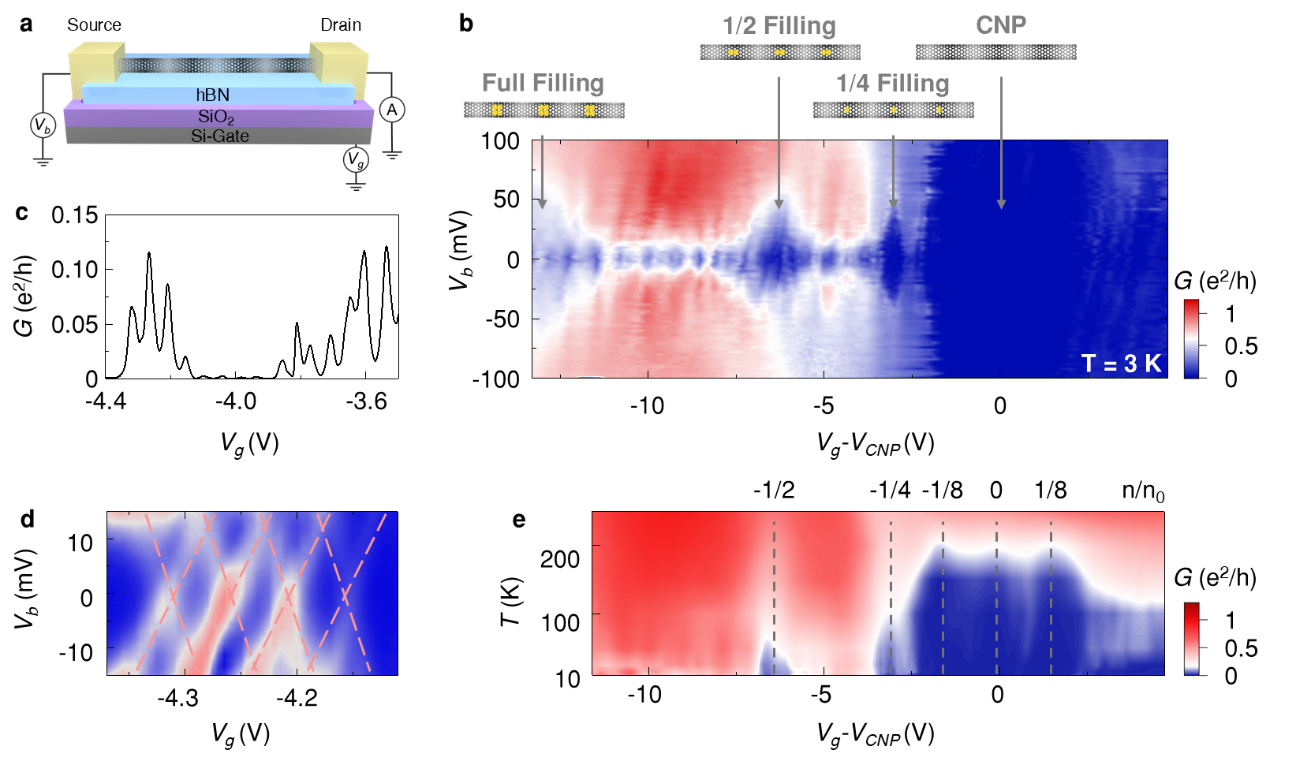}
\centering
\caption{\label{fig2} Emergence of correlated insulating states at 1/2 and 1/4 fillings. (a) Schematic of a CNT device with 1D moiré pattern. (b) Color plot of differential conductance as a function of $V_g$ and $V_b$ at 3 K. Insulating states can be observed at the charge neutrality point (CNP), the 1/4 filling, the 1/2 filling and full filling point of the hole mini-band. Insulating states at 1/4 and 1/2 filling are attributed to correlated insulators. (c) Gate-dependent differential conductance of the CNT device at 3 K. Equally spaced resistance peaks are induced by Coulomb blockade. The separation between these peaks $\Delta V_{\mathrm{g}}\approx\;57\;mV$ corresponds to the gate voltage required to charge the device with one electron/hole. (d) A zoom-in color plot of differential conductance as a function of $V_g$ and $V_b$. Well-defined Coulomb diamonds can be observed. (e) Color plot of conductance against carrier density and temperature. }
\end{figure*}

\section{correlated electronic phases in the 1D moiré superlattices}

To explore correlated electronic phases in the 1D moiré superlattices, we fabricated the as-grown CNT/hBN superlattices into field-effect devices and carried out transport measurement, the sketch of which is shown in Fig.\,2a. The channel length of the measured device is approximately 800 nm (See SI section 5), and the accurate moiré period of this device is 13.3 nm. Hence, there are only $\sim$\;60 supercells along the entire channel. To fully fill one miniband of the superlattice nanotube, we only need to supply $N_s\approx60\times4=240$ carriers, considering both spin and valley degeneracy. Therefore, we can readily tune the nanotube’s miniband to be half-filled and even fully-filled through an electrostatic gate. The device electrodes are made of Pd, a high work function metal, leading to a lower contact resistance on the hole-doping side\cite{29}. Therefore, in the following discussions, we mainly focus on the transport behaviors on the hole-doping side.

Figure 2b displays 2D mapping of the differential conductance $(dI/dV)$ as a function of both source-drain bias ($V_b$) and gate voltage ($V_g$) at a low-temperature of 3\;K. On the hole side, the highest conductance reaches $\sim1.2\,e^2/h$, indicating the excellent quality of the device. The most insulating region at $V_\mathrm{g}-V_\mathrm{CNP}=0\,\mathrm{V}$ corresponds to the fermi level being in the middle of the bandgap (the charge neutrality point, CNP). At low bias, a sequence of small diamonds of quantum oscillation appears due to the Coulomb blockade, since the nanotube channel can be considered as a quantum dot with a small capacitance. The Coulomb diamonds can be seen more clearly in a zoomed-in image in Fig.\,2d. Additionally, we plotted the device conductance as a function of gate voltage in Fig.\,2c, where the equally spaced peaks are direct evidence of the Coulomb blockade\cite{36,37,38}. The separation between these peaks $\Delta V_\mathrm{g}\approx\mathrm{\;57\;mV}$ corresponds to the gate voltage required to charge the device with one electron/hole.

In addition to these small Coulomb diamonds, several large diamond-like insulating regions also show up on the hole doping side, which is rather unusual. These large diamonds locate at $V_\mathrm{g}-V_\mathrm{CNP}=-3.4\;\mathrm{V}, -6.7\;\mathrm{V}, \mathrm{and}-13.5\;\mathrm{V}$ respectively. Noted that the ratio of these gate voltages 3.4 : 6.7 : 13.5 is approximately 1 : 2 : 4, and that $\Delta V_\mathrm{g}=3.4\mathrm{~V}$ is right the gate voltage required to fill one charge per supercell (1/4 filling, requiring $\sim$\;60 charges), as $57\mathrm{~mV}\times60=3.42\mathrm{~V}$ Therefore, we attribute these large diamonds to be 1/4 filling, 1/2 filling, and full filling of the valence miniband, respectively. The emergence of insulating states at these commensurate fillings is rather unusual and not expected in the single particle picture. Based on previous theories, we attribute the observed insulating states at 1/2 filling to be a correlated insulator caused by the on-site repulsion$(U)$. The insulating state at 1/4 filling requires more caution. One possible scenario is a CDW insulator (the CDW insulator phase can be viewed as a correlated insulator phase with unit cell doubled) caused by the repulsion within each supercell $(V)$ in the framework of the extended Hubbard model\cite{21}. Another possible scenario is likely a valley polarized phase where only one valley is half-filled. Importantly, all those possible insulators result from short-range interaction mediated by the superlattice (Umklapp scattering), which will be further discussed later. In addition, we also see signatures of insulating states at 1/8 filling (Fig.\,2e, and Figs.\,S11 and S12 in SI section 6), corresponding to filling one charge carrier for every two supercells, which could result from the Coulomb repulsion between nearest supercells, or it could be attributed to entering a pinned Wigner crystal phase\cite{39,40}. Evidence of the correlated insulating states at specific fillings was also observed in another CNT/hBN superlattice device (Fig.\,S13 in SI section 7).

Strikingly, even under conditions close to room temperature, the relative insulating resistance peaks at these commensurate fillings are still visible. Figure 2e shows a 2D color plot of the conductance against carrier density and temperature. As temperature increases, all the Coulomb\,-\,blockade\,-\,induced quantum oscillations fade away, but the increased resistance at 1/2 and 1/4 fillings is still visible up to 200 K (more data can be found in SI section 6), indicating rather strong electron-electron correlation (or Coulomb repulsions $U$ and $V$) in the 1D superlattices. The 1/8 filling state exhibits rather insulating over a wide temperature range, with almost no measurable current. Therefore, the subsequent discussions primarily revolve around the 1/2 and 1/4 fillings.

\section{Luttinger-liquid behavior of the 1D CNT superlattice at different fillings}

\begin{figure*}
\includegraphics[width=14cm]{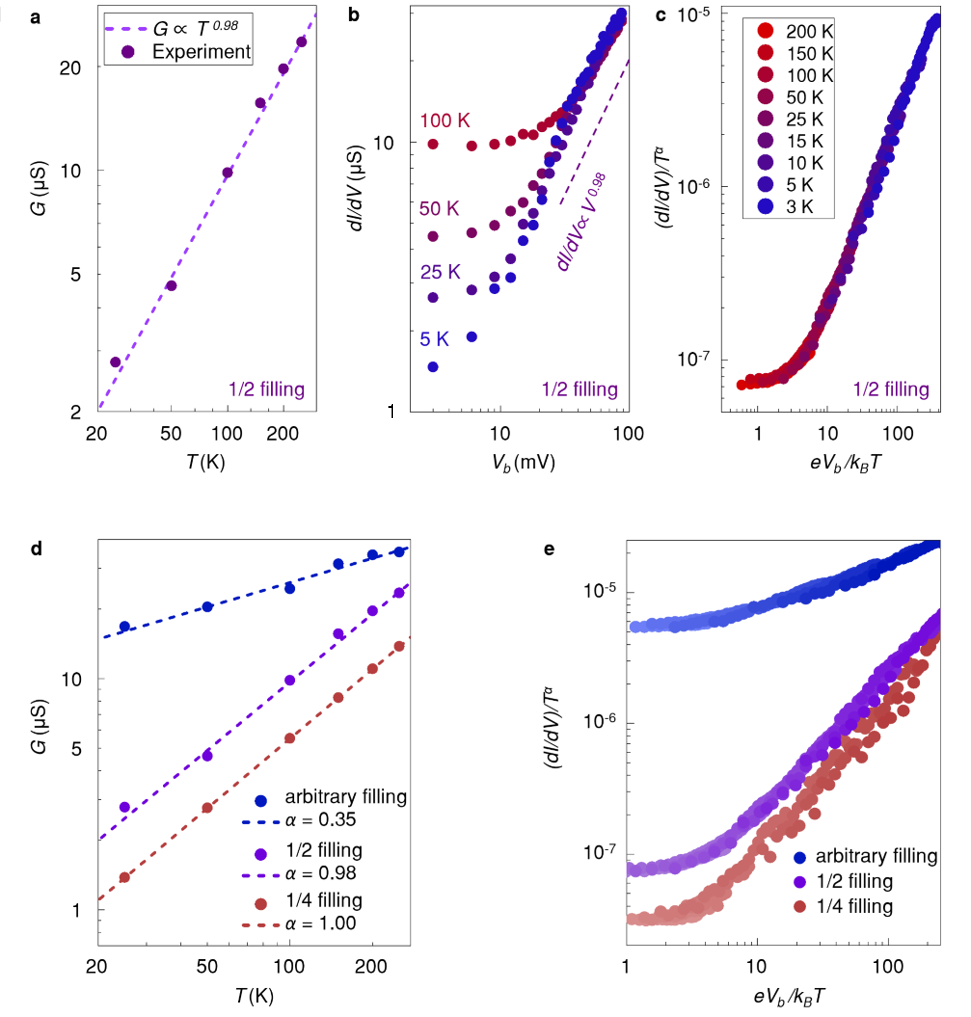}
\centering
\caption{\label{fig3}  Luttinger\,-\,liquid behavior of the 1D CNT superlattice at different fillings. (a) The conductance at 1/2 filling as a function of temperature T, showing a clear power law behavior with a component $\alpha\;{\sim}\;0.98$. (b) dI⁄dV measured at 1/2 filling as a function of electrical bias at a few representative temperatures. A power function fitting (dash line) yields the same power-law exponent. (c) The same data as in (b), but plotted as a scaled conductance $(dI/dV)/T^\alpha$ versus a scaled excitation $eV⁄(k_B T)$, where all data collapse to a single curve, as predicted by the Luttinger-liquid theory. (d) Temperature dependence of the conductance at three representative fillings (1/2, 1/4 and an arbitrary filling). (e) Scaled conductance versus scaled excitation at the three representative fillings, which contains all the data we measured in various temperatures ranging from 3 K to 200 K. All of them fit the universal scaling.}
\end{figure*}

To further investigate the electron correlation in the 1D superlattices, we employ the Luttinger\,-\,liquid theory. The correlated 1D electrons in carbon nanotubes can be fully characterized by the Luttinger parameter, g, which is defined as $g=[1+\frac{2U_C}{\Delta}]^{-\frac{1}{2}}$ and measures the strength of the electron correlation. Here $U_c$ is the charging energy of the CNT and $\Delta$ is the single particle level spacing\cite{27}. Non-interacting electron gases are described by $g=1$, while attractive interactions have $g>1$ and repulsive interactions have $g<1$. The Luttinger parameter $g$ can be experimentally determined through electron transport measurement. The tunneling of an electron into a Luttinger liquid follows a power law\cite{18} with a power exponent of $\alpha=(g^{-1}-1)/4$, due to the four conducting modes in CNTs and end-contacts. Previous studies have reported a typical value of $g\sim0.3$ for single-walled CNTs\cite{41,42}.

In the following, we study the Luttinger behavior of the 1D nanotube superlattice. We first focus on the Luttinger behavior at the 1/2 filling. Figure 3a displays conductance $G$ as a function of temperature $T$ on a double logarithmic scale, for the case of small bias $eV_b\ll k_BT$. The transport data show a clear power-law behavior $G(T)\propto T^\alpha$ with an exponent $\alpha=0.98$\;(approximately one, the dashed line), for $T$ ranging from 25 to 250 K. Note that for $\alpha=1$ the power-law dependence becomes a linear relationship, leading to a $T$-linear conductance. Figure 3b displays the differential conductance $dI/dV$ as a function of source-drain bias $V_b$ at a few representative temperatures, also plotted on a log-log scale. In the high-bias regime, $dI/dV$ increases as $V$ increases, and all curves at different temperatures fall onto a single curve with a trend that is well fitted by the same power-law exponent $\alpha$, namely, $dI/dV\propto V^\alpha$ (the dashed line in Fig.\,3b). The scaled conductance $\frac{dI}{dV}/T^\alpha$ versus the scaled bias $\frac{eV_{b}}{k_{B}T}$ with the same exponent $\alpha$ is plotted in Fig.\,3c, where the scaled conductance is constant as $\frac{eV_{b}}{k_{B}T}$ approaches zero and increases linearly when $\frac{eV_{b}}{k_{B}T}$ is significantly greater than one. All data points taken over one decade in $T$ and over two decades in $V_b$ collapse into a single curve, indicating a universal scaling behavior.

Similar scaling behavior is also observed at 1/4 filling and an arbitrary filling point, as displayed in Fig.\,3e and 3f. At the 1/4 filling, fitting the experimental data with the power law yields $\alpha\sim1.00$, also indicating a T-linear conductance. At the arbitrary filling point, a much smaller $\alpha\sim0.35$ and a larger $g\sim0.42$ is extracted (Fig.\,3d), which is comparable to the values reported in previous studies for intrinsic CNTs\cite{41}. As seen in the scaled log-log plot (Fig.\,3e),  $\frac{dI}{dV}/T^\alpha$ versus $\frac{eV_{b}}{k_{B}T}$, all scaled data points under the same gate voltage collapse into a single curve. For low scaled-bias $\frac{eV_{b}}{k_{B}T}<3$, the scaled conductance develops plateaus whose value depends only on the filling level; for high bias $\frac{eV_{b}}{k_{B}T}>3$, the differential conductance increases with the bias voltage as $dI/dV\propto V^\alpha$, where the power law index $\alpha$ is also filling dependent. 

\begin{figure*}
\includegraphics[width=17cm]{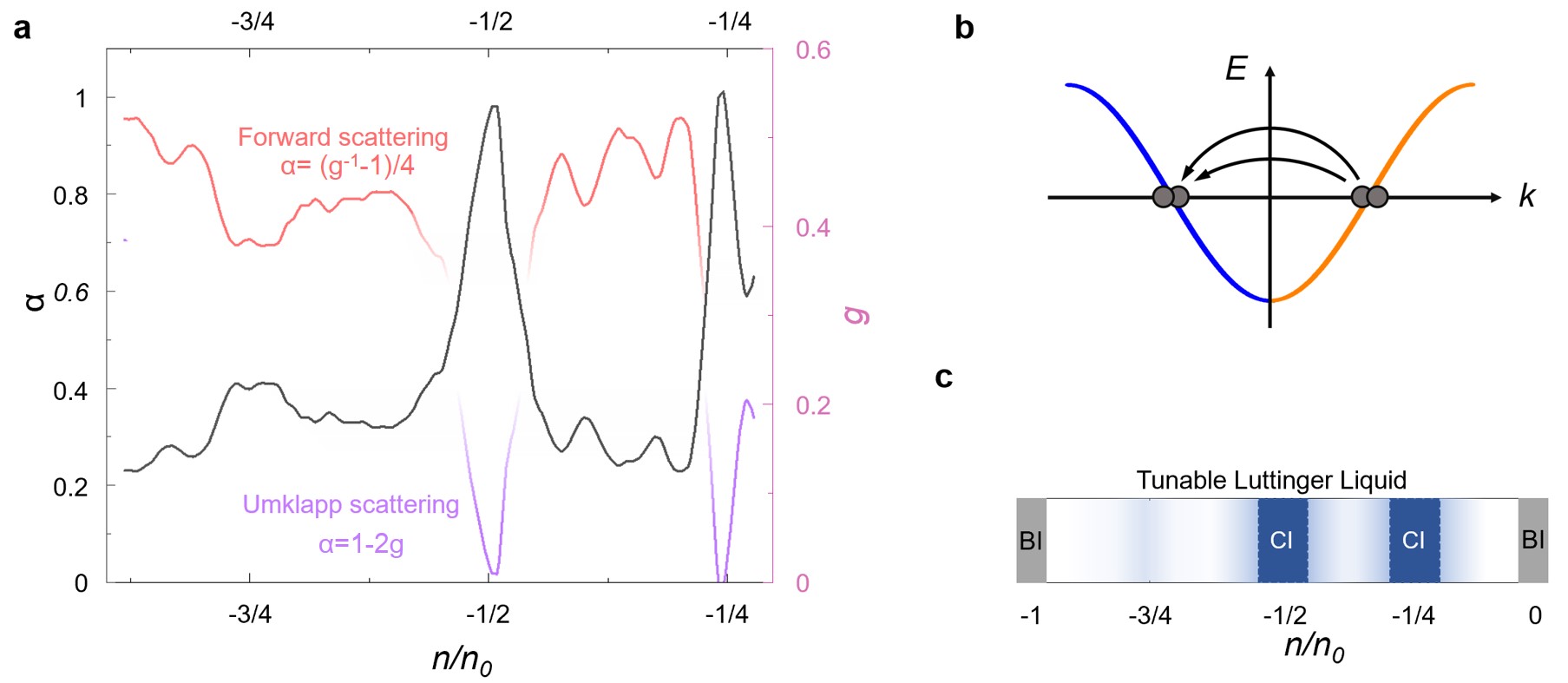}
\centering
\caption{\label{fig4}  Filling\,-\,dependent Luttinger parameter and phase diagram of the 1D CNT moiré superlattice. (a) Extracted power-law exponent $\alpha$ (black line) and Luttinger parameter g (red and purple lines in different filling regions) as a function of the filling factor. The Luttinger parameter g drops to zero at 1/4 and 1/2 fillings, indicating possible breakdown of the Luttinger liquid. (b) Schematic of electron\,-\,electron Umklapp scattering in one\,-\,dimensional moiré superlattice at half filling. (c) A schematic phase diagram for the 1D CNT superlattice, which behaves as a tunable Luttinger liquid. The correlated insulators appear at 1/4 and 1/2 filling. Gray areas near zero filling and full filling refer to band insulators (BI).}
\end{figure*}

We further extract the power law index $\alpha$ in a wide range of filling levels and plot in Fig.\,4a (black curve). Two prominent peaks appear at 1/2 and 1/4 fillings, and the peak value approaches $\sim$\;1. When filling is away from 1/2 or 1/4, the power law index quickly drops to below 0.5, indicating a significant difference between the commensurate (1/2 and 1/4) filling and other fillings.

At last, we investigated the Luttinger parameter $g$ of the 1D superlattice at different fillings. For general fillings, using $\alpha=(g^{-1}-1)/4$, we obtained $g$ from experimentally measured $\alpha$ and plotted it as the red line in Fig.\,4a. The value of $g$ ranges between 0.3 $\sim$ 0.5, comparable to the values from previous studies\cite{41,43,44}. However, there are strongly suppression of g around 1/4 and 1/2 fillings. Considering that $g=[1+\frac{2U_c}{\Delta}]^{-\frac{1}{2}}$, suppression of g around 1/4 and 1/2 fillings indicating enhanced interaction around those special fillings. More interestingly, it is predicated that Umklapp scattering becomes relevant at 1/2 filling\cite{18}. In the Umklapp process, two forward moving electrons at the Fermi surface can both be scattered to moving backwards (Fig.\,4b). As a result, a metallic nanotube transits into a correlated insulator, which significantly modifies the conductance-temperature power-law relations to be $G(T)\;{\sim}\;T^{1-2g}$. If we use the Umklapp process modified formula of $\alpha=1-2g$, a newly extracted value of $g$ is plotted as purple line in Fig. 4a. Notably, at 1/2 and 1/4 filling,  $g$ drops to its theoretical minimum, zero, considering that $g=[1+\frac{2U_c}{\Delta}]^{-\frac{1}{2}}\geqslant0$. This corresponds to the case $U_c\gg\Delta$, meaning the system is at the strongly interacting limit where electrons are frozen. This indicates possible breakdown of the Luttinger liquid theory.

Figure 4c displays a schematic phase diagram for the 1D CNT superlattice. Basically, the 1D moiré superlattice behaves as a tunable Luttinger liquid, with a filling-dependent Luttinger parameter $g$ ranging from 0.3 to 0.5 at general fillings. At low temperature, the correlated insulating phases appear at 1/4 and 1/2 fillings. Right above the correlated insulating states is the $T$-linear conductance states with strongly suppressed Luttinger parameters.

\section{Conclusion}

In summary, we carried out transport measurements on a directly grown 1D moiré superlattice of CNTs on hBN. Thanks to the large moiré period, we are able to adjust the dominant electron-electron interaction by changing the gate voltage, transiting from forward scattering at generic filling to Umklapp scattering at the commensurate 1/4 and 1/2 filling, leading to a widely tunable Luttinger liquid. At 1/2 and 1/4 fillings, we observed correlated insulating states, which can be explained in the framework of the extended Hubbard model\cite{21}. We also observe a T-linear conductance at these commensurate fillings over a range of temperatures, which leads to a strongly suppressed Luttinger parameters, indicating a state with extreme correlation. In addition to the correlated insulator, Luttinger liquid under periodic potential fields also has a rich phase diagram\cite{21,45}. Therefore, the 1D CNT/hBN superlattice provides an ideal experimental platform to study (quasi) 1D correlated phases. For example, by coupling two moiré Luttinger chains, one can realize the Luther-Emery liquid phase\cite{46}. 

\section{ACKNOWLEDGMENTS}
This work is supported by the National Key $R\&D$ Program of China (No.\,2021YFA1202902, No.\,2022YFA1402702, No.\,2021YFA1401400, and No.\,2022YFA1405400), the National Natural Science Foundation of China (No.\,12074244, No.\,12274290, No.\,12274289, No.\,12174249, and No.\,92265102), and the open research fund of Songshan Lake Materials Laboratory (No.\,2021SLABFK07). Z.S. acknowledges support from SJTU (21X010200846), and additional support from the Shanghai talent program. X.Y.X. acknowledges support from Shanghai Pujiang Program under Grant No.\,21PJ1407200, Yangyang Development Fund, and startup funds from SJTU. M.Q. acknowledges the support from the Innovation Program for Quantum Science and Technology (under grant no.\,2021ZD0301900) and the Yangyang Development Fund. K.W. and T.T. acknowledge support from the JSPS KAKENHI (Grant Numbers 20H00354 and 23H02052) and World Premier International Research Center Initiative (WPI), MEXT, Japan.

\section{Appendix:Methods}
\subsection{\label{sec:level2}CNT/hBN heterostructure growth.}
First, hBN flakes were mechanically exfoliated onto SiO$_2$/Si substrates, followed by exposing to hydrogen plasma at $\mathrm{300~^\circ C}$ to remove all organic residuals. Then, 0.05nm-thick Fe film was deposited on the hBN-covered SiO$_2$/Si surfaces through thermal evaporation (evaporation rate:\,$\sim\;0.005\;nm/s$, base vacuum pressure:$\sim1\times10^{-6}$mbar), serving as catalytic nanoparticles for the CNT growth. After that, the chips were put into a 1-inch-diameter quartz tube furnace (Anhui BEQ Equipment Technology) and gradually heated up to the growth temperature ($\mathrm{800-850~^\circ C}$) under hydrogen and argon gas mixture at atmospheric pressure. When growth temperature was reached, argon was replaced by methane to commence the CNT growth. After a typical growth duration of 60 min, the system was cooled down to room temperature under a protective hydrogen and argon atmosphere. 

\subsection{\label{sec:level2}Atomic force microscopy (AFM) characterization.}
A commercial AFM (Cypher S, Asylum Research, Oxford Instruments) was used to image the topography of the as-grown CNT samples. High resolution topography images were scanned in AC mode (tapping mode) with probes of models FS-1500 and PFQNE-AL. For the lattice resolution imaging of hBN, friction mode and probes with model SCM-PIT-75 were used.
 
\subsection{\label{sec:level2}MD simulations of the stacking energy.}
The simulated model system consists of a 100-nm-long armchair CNT with chiral indices (7, 7), and a fixed hBN substrate. The intra-layer interactions within the CNTs and the hBN substrate were computed via the second generation of REBO potential\cite{47} and the Tersoff potential\cite{48}, respectively. The interlayer interactions between the CNT and the hBN substrate were described via the registry\,-\,dependent ILP\cite{49} with refined parametrization\cite{50}, which we implemented in LAMMPS\cite{51}. At each fixed twist angle, we relax the entire system to obtain the structure with the lowest energy.

\subsection{\label{sec:level2}Device fabrication and electrical characterization.}
Devices of the CNTs are fabricated using standard nanolithography techniques. First, a combination of E-beam lithography (EBL) and oxygen plasma was used to etch out all other CNTs, with only the target superlattice nanotube left for the transport measurement. Then, another e-beam lithography defined areas for the metal contacts (5\;nm\;Pd\,/\,70\;nm\;Au). Transport measurements were performed in a cryostation (Model: s50, Montana Instruments) with a base temperature of 2.7 K. We applied standard lock-in technique to measure the devices resistance, with 37.7 Hz excitation frequency and less than 200 $\mu V$ alternating excitation bias voltage, achieved by using a 1/1000 voltage divider.
 
\bibliography{apssamp}

\end{document}